\documentclass[amsmath,amssymb,floatfix,reprint,twocolumn,superscriptaddress,prb]{revtex4-2}

\usepackage{graphicx}
\usepackage{epsfig}
\usepackage{hyperref}
\usepackage{xcolor}
\usepackage{bm}

\begin{document}

\title{$1/f^{3/2}$ Spectral Density at the Phonon Bottleneck}
\author{Steven T. Bramwell}
\date{\today} 

\affiliation{
London Centre for Nanotechnology and Department of Physics \& Astronomy, University College London,
17-19 Gordon Street, London, WC1H 0AJ, United Kingdom}

\begin{abstract}
The common observation of anomalous `$1/f^\alpha$' relaxation with $\alpha<2$ constitutes one of the enduring mysteries of condensed matter physics. Here it is shown that a $1/f^{\alpha}$ spectral density, with $\alpha = 3/2$, can arise in the response of an ensemble of two--level systems coupled to a heat bath by means of a system of Bosonic quasiparticles. The model considered is the classic model of Faughnan and Strandberg of the phonon bottleneck, and the anomalous response is associated with an approximate non-equilibrium steady state of the phonons maintained by slow spin relaxation. The frequency dependence of the response to an applied field is calculated analytically, revealing the emergence, in the limit of a strong bottleneck, of $\alpha=3/2$ behaviour over a diverging range of frequencies. The application of this result to experimental systems is discussed and comparisons are drawn with other systems that exhibit anomalous relaxation. 
  \end{abstract}

\maketitle

\section{Introduction}\label{Intro}

Exponential, or Debye, relaxation has a spectral density that tends to $1/f^2$ at high frequency $f$. However many real condensed matter systems -- conducting~\cite{DuttaHorn,Weissman}, dielectric~\cite{Jonscher, Phillips, NonDebye} and magnetic~\cite{Gunnarsson, Ocio, Sangregorio, Thomas, Hallen, Billington} -- show power law behavior $\sim1/f^\alpha$ with an exponent $\alpha$ that lies in the range $1\le\alpha < 2$. While a number of particular mechanisms for this behavior have been established (see e.g. Refs.~\cite{Gunnarsson, Ocio, Sangregorio, Thomas, Hallen} for the magnetic case), in general it remains a mystery as to why $1/f^\alpha$ behavior is so widespread. A common explanation~\cite{DuttaHorn} is that the power law is constructed from a wide range of exponential timescales, but in physical terms the range of timescales required (many decades) is usually unrealistic. An exception might be where the time scales are associated with an underlying broad range of length scales, as occurs in a critical or fractal system, but such instances cannot account for the majority of observations. This leads one to suspect that there are more fundamental causes at play, arising in non-equilibrium physics and quantum dynamics. In this context, simple and realistic models and experiments that produce well-defined exponents $\alpha$ are of evident interest to signpost the way to fundamental causes. 

Straightforward mathematics, and more detailed studies of quantum systems~\cite{Fonda}, reveal two very general requirements for non--exponential relaxation. First, the system needs `memory' of its relaxation, and second, memory is enhanced by disconnection of the system from its environment. In magnetism, at least partial disconnection from an environmental heat bath is always a factor to consider at sufficiently low temperature, because that connection is tyoically mediated by a rapidly decreasing population of increasingly long-lived phonons. This gives rise to the `phonon bottleneck' which was characterised many years ago in dilute spin systems~\cite{Strandberg,FS,SJ,Standley,Stoneham, Brya,GillOrb,Gill,Roinel}. Unsurprisingly, the phonon bottleneck is associated with non-exponential relaxation~\cite{FS,Standley}, but it is germane to ask the question, is it associated with any of the common analytic forms of non-exponential relaxation?  Rather surprisingly, this question never seems to have been addressed in the literature. 

\begin{figure}[!ht]
	\centering
	\includegraphics[width=0.9\linewidth]{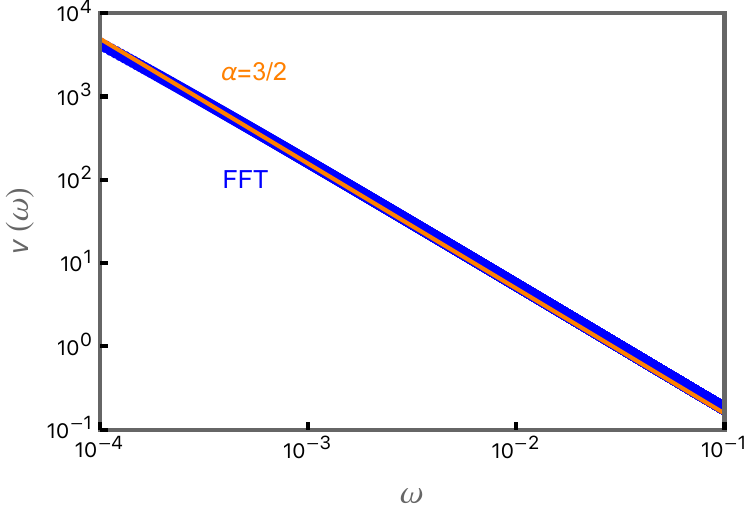}
	\caption{Phonon bottleneck relaxation according to the model of ref. \cite{FS} with parameters $a=100, \sigma = 500$ (parameters defined in Section \ref{Review}). Fast Fourier transform (FFT) of the numerically evaluated relaxation function $v(|t|)$ (blue dots) compared with a line (orange) of slope $3/2$ on a log-log scale, to illustrate an extensive range of $\sim 1/\omega^{3/2}$ relaxation. } \label{Fig1}
\end{figure}

The aim of this paper, therefore, is to answer the question stated above, with respect to the canonical model of the `phonon bottleneck' due to Faughnan and Strandberg~\cite{FS}. The question is simple, and an answer is immediately suggested by numerically solving the equations of Ref.\cite{FS} in the limit of a `strong' bottleneck. The ensuing relaxation function, when digitally (`Fast') Fourier transformed, shows a clear $1/\omega^{3/2}$ dependence, as shown in Fig. \ref{Fig1} (here $\omega = 2 \pi f $)). The rest of the paper is concerned with understanding this result in terms of an analytic solution, exploring its relevance to real systems and discussing its implications for non-Debye relaxation in general.  

In the following, the phonon bottleneck is reviewed (section \ref{Review}), the analytic result derived (section \ref{Analytic}), relation to real systems discussed (section \ref{Application}) and conclusions drawn (section \ref{Conclusions}).

\section{Review of the phonon bottleneck}\label{Review}

It is well established that isolated magnetic moments (`spins') relax through three main processes: the direct process, the Raman process and the Orbach process~\cite{SJ,Stoneham,Gill}. The latter two processes tend to die away rapidly at low temperature leaving only the direct process, by which a phonon of appropriate energy is absorbed or emitted every time a spin flips. The spin-phonon interaction arises from a combination of lattice vibrations modulating the crystal electric field, and spin-orbit coupling in the case of electronic spins, or the nuclear quadrupole interaction in the case of nculear spins. In a perturbation treatment, matrix elements depend on phonon occupation numbers which become natural variables in the problem, alongside spin states (`up' or `down' in the simplest case)~\cite{Strandberg}. If the magnetic heat capacity is much larger than the phonon heat capacity, which it typically is at low temperature, and if the phonon relaxation time starts to exceed the spin-lattice relaxation time, then magnetic relaxation drives phonon populations out of equilibrium, and magnetic relaxation is then determined by the rate of transfer of energy from the phonons to the bath. In the simplest case this manifests by a change in temperature dependence of the relaxation rate from $\sim T$ to $\sim T^2$~\cite{SJ} (here $T$ is the bath temperature). The process can be treated approximately as a three-temperature problem (temperatures of spin system, phonons and bath)~\cite{Stoneham}, but more generally the phonon and spin temperatures become poorly defined concepts, and it is more meaningful to work directly with phonon and spin state populations~\cite{Strandberg}. 

The canonical model of the phonon bottleneck due to Faughnan and Strandberg~\cite{FS} makes two basic assumptions. The first is
\begin{equation}
\frac{dN_i}{dt} = 
-\frac{dN_j}{dt} = 
-\rho A \left[N_i (p +1) - N_j p \right]
\end{equation}
where
$N_i$ as the number of spins in the excited state, $N_j$ is the number of spins in the ground state, $p$ is the phonon occupation number in a narrow band at the frequency of the transition,  $\rho$ is the number of phonon states, $A$ is a rate constant for stimulated emission or absorption, and the factor `1' represents spontaneous emission. This equation may be manipulated into the form given in Ref. \cite{FS} using the additional condition that $N^i+N^j = N$, the number of spins. The second assumption is: 
\begin{equation}\label{pp}
\frac{d(p-p^0)}{dt} = \frac{-1}{2\rho \Delta \nu} \frac{d( N_i-N_j)}{dt} - \frac{(p-p^0)}{\tau_{\rm ph}}.  
\end{equation}
where (here and in what follows) the superscript $0$ indicates equilibrium values, $\Delta \nu$ is the relevant band width and  $T_{\rm ph}$ a phonon relaxation time that describes absorption by the heat bath or scattering into other phonon modes. This leads to two coupled equations of motion for the spin and phonon populations, which are conveniently rewritten in terms of dimensionless properties as follows. 
\begin{equation}\label{v}
\frac{d v}{d t} = -v +\sigma u (1-v)
\end{equation}
\begin{equation}\label{u}
\frac{d u}{d t} = a\left[(v - u)- \sigma u (1-v) \right]
\end{equation}
where
\begin{equation}\label{udef}
u= \frac{1}{\sigma} \frac{p-p^0}{p^0+\frac{1}{2}}
\end{equation}
and the variable $v$ is defined as
\begin{equation}\label{vdef}
v = \frac{(N_i^0-N_j^0) - (N_i - N_j)}{(N_i^0 - N_j^0)}. 
\end{equation} 
Here the parameter $a= \tau_{\rm sl}/\tau_{\rm ph}$ where 
$\tau_{\rm sl}$ is the spin-lattice relaxation time and the parameter $\sigma = b/a$. The latter can also be cast as: 
\begin{equation}\label{sigdef}
\sigma = \frac{c_s/\tau_s}{c_p/\tau_p}. 
\end{equation}
where the $c$'s are specific heats. A large value of $\sigma$ indicates that the energy exchange rate between spins and phonons greatly exceeds that between phonons and bath -- hence it is a measure of the strength of the `phonon bottleneck'. Also it should be noted that the time $t$ here is measured in units of $\tau_{\rm sl}$.  

The limiting situation of interest arises where $\sigma, a \gg 1$. This is the scenario illustrated in Fig. \ref{Fig1}, which shows the numerical solution of the coupled equations for $a=100, \sigma = 500$.  Under such conditions the phonons are strongly bottlenecked but the spins relax relatively slowly, continually maintaining a far from equilibrium phonon population. This leads to an approximate steady state where the rate of change of the phonon population can be neglected in Eq. \ref{pp}. Thus $|du/dt| \ll a$, and the right hand side of Eq. \ref{u} may be set to zero. This reduces Eqs. \ref{v}, \ref{u} to a single equation depending only on $\sigma$: 
\begin{equation}\label{reducevu}
\frac{d v}{d t} = \frac{-v}{1+\sigma(1-v)}= -u.
\end{equation}
This equation has an exact solution for time $t$:
\begin{equation}\label{time}
t= -\sigma \left(1-v\right) - (1+\sigma)\ln \left(v\right). 
\end{equation}

Faughnan and Strandberg~\cite{FS} showed that numerical solutions of both the coupled equations and the limiting equation Eq. \ref{reducevu} gave non-exponential relaxation, but the functional form of the relaxation was not explored. In addition they showed that Eq. \ref{time} was a close approximation to the coupled equations for realistic parameters. In later work, Standley and Wright~\cite{Standley} found excellent agreement of these equations with experiment, as discussed further below. Later work on the phonon bottleneck largely developed the approach of Ref.\cite{FS} to include other factors, which include phonon diffusion~\cite{Stoneham}, negative temperatures~\cite{Brya}, and the more complex adaption of the model to describe the Orbach processes~\cite{GillOrb, Gill}. 

\section{Analytic form of the relaxation}\label{Analytic}

In order to establish notation and the meaning of relevant quantities, it is useful to briefly review some concepts of linear response theory. In linear response one typically imagines a system at equilibrium in an applied field $H$, such that the magnetization is $M^0 = M(T \le 0) =  \chi_T H$. The field is then removed such that the system relaxes according to $M(t\ge 0) = R(t)$. Here $R(t)$ is a relaxation function, formally an {\it even} function of $t$. Clearly $R(0) = \chi_T$ and one can define a normalised function, $F(t) = R(T)/\chi_T$, whose Fourier transform is the spectral shape function $F(\omega)$. The response function $\chi(t)$ is defined by the convolution 
\begin{equation}
M(t) = \int_{-\infty}^t H(t') \chi(t-t') \, dt'
\end{equation}
and it follows that $\chi(t <0) = 0$ and $\chi(t \ge 0 = -d R(t)/dt)$. The Fourier transform of $\chi(t)$ is the complex susceptibility $\chi(\omega) = \chi'(\omega) + i \chi''(\omega)$, which connects the frequency dependent magnetization and field,  $M(\omega) = \chi(\omega) H(\omega)$. It is possible to show that~\cite{power}
\begin{equation}
\frac{2|\chi''|}{\chi_T\omega} = F(\omega).  
\end{equation}
A time varying field of particular importance is $H(t) = h e^{i \omega t}$. Then, if the reponse is $M(t) = m e^{i\omega t}$, the ratio $\chi=m/h$ is equal to $\chi(\omega)$ derived from the Fourier transform of the response function. Explictly, the relationship is: 
\begin{equation}
\chi(\omega) = \int_{0}^{\infty} \chi(t) e^{-i\omega t}\, dt= \frac{m}{h}.  
\end{equation}

The model of Faughnan and Strandberg (section \ref{Review}) may now be compared to these relationships. If $i$ denotes spin up, and $j$ denotes spin down, then the magnetic moment $\mathcal{M} = N_i-N_j$ (i.e opposite to the spin direction but this sign choice is ultimately not important) and the equilibrium magnetic moment is $\mathcal{M}^0 = N_i^0-N_j^0$. It follows that
$
v=(\mathcal{M}^0-\mathcal{M})/\mathcal{M}^0,
$
a normalised measure of the deviation of magnetic moment, or magnetization, from its equilibrium value. For example, if a field is suddenly applied to an equlibrium unmagnetized system, then $v$ will evolve from $v=1$ to $v=0$ as the new equilibrium is established at $\mathcal{M}= \mathcal{M}^0$.  

The alternative scenario of removing a field here leads to conceptual difficulties in physical terms. These are discussed in detail below, but for the time being one may proceed in a purely mathematical sense. Then, from the above definitions, it can be shown that $v(t\ge 0) = F(t)$, so the symmetrised $v(|t|) = F(t)$. In fact the limiting Eq. \ref{time} can be inverted to give $v(t)$, and hence $F(t)$ exactly:
\begin{equation}\label{Lambert}
   v(|t|)=F(t) = -\frac{1}{1+\sigma} W\left(-\frac{\sigma  \left(e^{-\sigma -|t|}\right)^{\frac{1}{1+\sigma}}}{1+\sigma}\right),
\end{equation}
where $W$ is the Lambert W-function. 

The function Eq. \ref{Lambert} does not have a known Fourier transform but it can be integrated, giving 
\begin{equation}\label{Fzero}
F(\omega = 0) = \sigma+2,  
\end{equation}
a result that will be used below. 

An analytic expression for $\chi(\omega)$ can now be derived. Thus Eq.  \ref{reducevu} may be rewritten as a function of the relative magnetization $M/M^0$, such that 
\begin{equation}\label{meqreduced}
1 - \frac{M}{M^0} = \frac{1}{M^0}\frac{dM}{dt}\left(1 + \sigma \frac{M}{M^0}\right),
\end{equation}
where $M^0=H \chi_T$. Now consider the case where $ H = h e^{i\omega t}$ and $M = m e^{i\omega t}$. It follows that: 
\begin{equation}\label{chieq1}
 \chi_T - \chi = i \omega \chi \left(
 1+ \sigma \frac{\chi}{\chi_T}, 
 \right),
\end{equation}
from which 
\begin{equation}
\frac{\chi}{\chi_T}  =  \frac{i \left(-\sqrt{4 i \sigma \omega -\omega^2+2 i \omega+1}+i w+1\right)}{2 \sigma \omega}
\end{equation}
consisting of the real and imaginary parts:
\begin{equation}\label{chip}
\frac{\chi'}{\chi_T} =
\frac{-\omega + (1 + (2 + 16 \sigma (1 + \sigma)) \omega^2 + \omega^4)^{1/4}
   \sin\phi}{2 \sigma w}
\end{equation}
\begin{equation}\label{chipp}
\frac{\chi''}{\chi_T} =
\frac{1-(1 + (2 + 16 \sigma (1 + \sigma)) \omega^2 + \omega^4)^{1/4}
   \cos\phi}{2 \sigma w}
\end{equation}
where $\phi = \frac{1}{2} \arg[1 + i (2 + 4 \sigma + i \omega) w]$ and $ \cos\phi, \sin \phi \rightarrow \frac{1}{\sqrt{2}}$ for large $\sigma$.

\begin{figure}[!ht]
	\centering
	\includegraphics[width=0.9\linewidth]{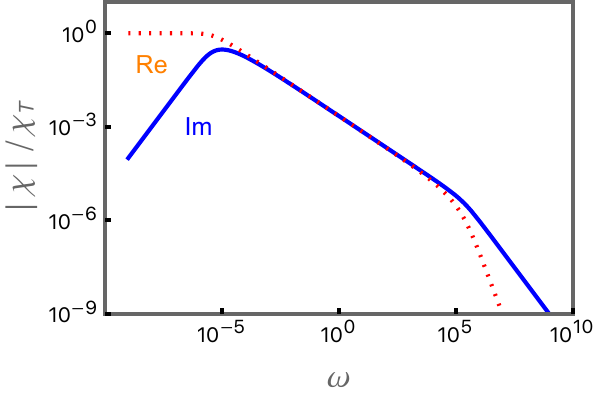}
    \includegraphics[width=0.9\linewidth]{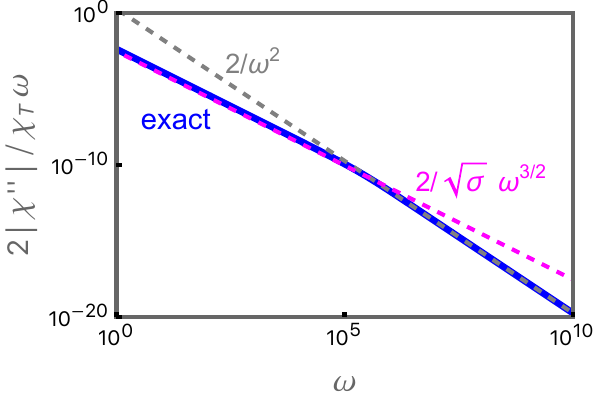}
	\caption{Frequency dependence of the derived complex susceptibility for $\sigma= 500$. Upper plot: real part (orange dotted line, Eq. \ref{chip}) and imaginary part (full blue line, Eq. \ref{chipp}) of the the analytic complex susceptibilty, showing three distinct frequency regimes. Lower plot: the corresponding function $2|\chi''|/\chi_T\omega$ in the region of the high frequency `crossover', comparing the `exact' Eq. \ref{chipp} (blue line), the intermediate-frequency approximation Eq. \ref{asymp} (orange dashed line) and the high frequency approximation $2/\omega^2$ (gray dashed line). } \label{Fig2}
\end{figure}
The real and imaginary parts of the susceptibility, on log-log scales, are illustrated in Fig. \ref{Fig2} where there is a very clear structure of three linear regimes. According to Eq. \ref{chipp}, for large for large $\omega < \sigma$, 
\begin{equation}\label{asymp}
\frac{2|\chi''|}{\omega\chi_T} \rightarrow \frac{2}{\sqrt{\sigma} \,
\omega^{3/2}},  
\end{equation}
which picks out the `middle' regime in Fig. \ref{Fig2}, while for $\omega>\sigma$, high order terms in Eq. \ref{chipp} ensure that the true asymptotic behavior is $1/\omega^2$, with a crossover to this form at $\omega \sim \sigma$. Thus `$\alpha = 3/2$' scaling applies to the range $1/\sigma <\omega < \sigma $ and only the divergence of $\sigma$ would extend its range to infinity. Nevertheless, as illustrated in Fig. \ref{Fig2}, for finite $\sigma$, the crossover is particularly sharp and has negligible effect on the power law scaling at smaller $\omega$.

The function $\chi(\omega)$ thus calculated gives the exact steady state response to an applied oscillating field. The function $2|\chi''|/\chi_T\omega$ unsurprisingly coincides with a numerical Fourier transform of the exact $F(t)$ at high frequency, but, somehwat unexpectedly, as $\omega\rightarrow 0$ it tends to the value $2\sigma +2$, in contrast to $\sigma+2$ given by Eq. \ref{Fzero}. This perhaps suggests, for finite $\sigma$, there may not be a single response function that simultaneously covers both extreme cases of a step function applied field and a sinusoidal applied field. 

\begin{figure}[!ht]
	\centering
	\includegraphics[width=0.9\linewidth]{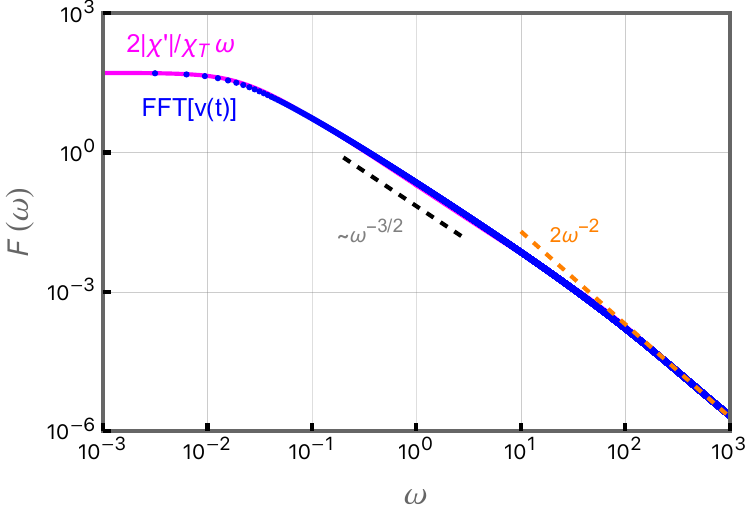}
	\caption{Phonon bottleneck relaxation in the limiting case with bottleneck parameter $\sigma = 50$. Fast Fourier transform (FFT) of the exact relaxation function $v(|t|)$ (blue dots, Eq. \ref{Lambert}) compared with the analytic $2|\chi''|/\chi_T\omega$ (magenta line, Eq. \ref{chipp}), showing distinct regimes of $\sim1/\omega^0, \sim 1/\omega^{3/2}$ and $2/\omega^2$ (Eq. \ref{asymp}) behaviour. In the analytic curve, the value of $\sigma$ is rescaled by a factor of $2$ to exactly match $v(\omega = 0)$. With this rescaling, and no fitting, the two curves are almost coincident at all frequencies, illustrating a rare instance where a `$1/f^{\alpha}$' spectral density (with $\alpha = 3/2$) can be calculated analytically. } \label{Fig3}
\end{figure}

Nevertheless, if $2|\chi''|/\chi_T\omega$ is adjusted by rescaling $\sigma\rightarrow \sigma/2$, such that it exactly coincides with the expected $F(\omega = 0)$, then, as shown in Fig. \ref{Fig3} (which includes no fitted parameters), it provides an astonishingly close approximation to the numerical Fourier Transform of the exact $F(t)$ over the whole frequency range. 
From this one may conclude that the true Fourier transform of $v(|t|) = F(t)$ also shows the emergence of a `$1/\omega^{3/2}$' regime, that will cover the whole domain of $\omega$ in the limit $\omega \rightarrow \infty$. 

The three-slope function in frequency space may be anticipated (at least with the benefit of hindsight!) from Eq. \ref{reducevu}. Thus, if $v$ can be approximated to either 1 (short times) or 0 (long times) in the denominator, then exponential relaxation is obtained, and this clearly translates through to the frequency response at large and small frequencies respectively. Also, the $1/\omega^{3/2}$ spectral function in frequency translates to $t^{-1/2}$ relaxation in the time domain and there is a hint of this in Eq. \ref{reducevu}.
Thus, at short times $v\rightarrow1$, but for sufficiently large $\sigma$, there appears to be a region at $1-v \gg 1/\sigma$ where this equation approximates:  \begin{equation}\label{approx1}
v \approx 1-\sqrt{2t/\sigma} \approx e^{-\sqrt{2t/\sigma}}. 
\end{equation}
Despite these hints, the different regions of the relaxation curve are much less clearly distinguished in the time domain than they are in frequency space, and this likely explains why a `$\sqrt{t}$' dependence has not been noticed, or considered, before.

\section{Application to real systems}\label{Application}

In applying this result to real systems, three cases must be clearly distinguished. 

The first case is where $2|\chi|/\chi_T\omega$ simply represents the Fourier transform of the normalised relaxation function $v(|t|)$. If experimental relaxation data is consistent with the $F(t)$ of Eq. \ref{Lambert} then it is most likely consistent with its Fourier transform. 
There are at least two studies, by Standley and Wright~\cite{Standley}, and by Roinel \textit{et al.}~\cite{Roinel}, that tested Eq. \ref{time} and hence Eq. \ref{Lambert}. Both studies reported excellent agreement with the experimental data. 
It was, of course, necessary to introduce units of time, by writing $t = \mathcal{W} t'$  where $t'$ is the dimensional time. Standley and Wright used esr to study the magnetic relaxation of an organic copper complex (copper dipyrromethene), following a microwave pulse, fitting their data at 2.45 K (for example) with $\mathcal{W}=1.06/{\rm ms}$, $\sigma = 1.57$.  Roinel \textit{et al.} used nmr to measure the relaxation of the polarisation (magnetization) of $^{169}$Tm nuclear spins in TmPO$_4$ at $T=50$ mK,  $H_1=1$ T, fitting their data with $\mathcal{W} = 1/(26~{\rm hr})$ and $\sigma = 17$. 

\begin{figure}[!ht]
	\centering
	\includegraphics[width=0.9\linewidth]{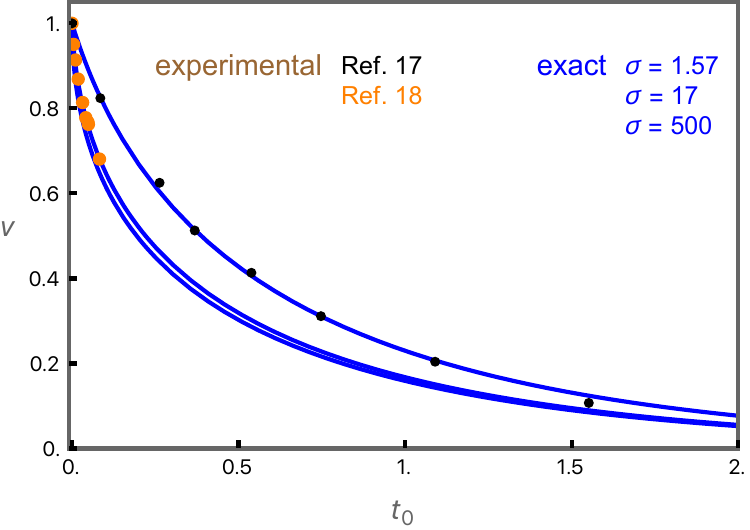}
    \includegraphics[width=0.9\linewidth]{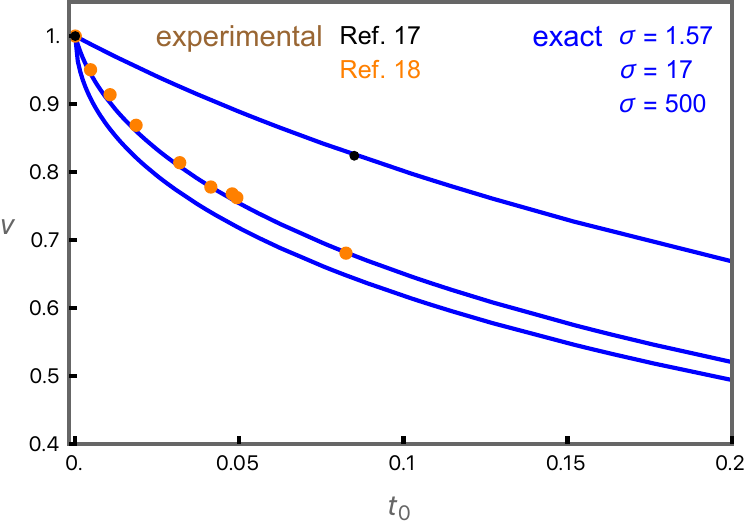}
	\caption{Replotted experimental data~\cite{Standley, Roinel} on a reduced timescale $t_0$ defined in the text. The lower figure is an enlargement of the upper one for short times. The black points are the data of Ref. \cite{Standley}, and the orange points are the data of Ref. \cite{Roinel}. The blue lines are the exact solution, Eq. \ref{Lambert} for (top to bottom) $\sigma = 1.57$~\cite{Standley}, $\sigma = 17$~\cite{Roinel} and $\sigma = 500$, a large value for comparison. The figure illustrates how closely experimental systems approach the exact limiting form, Eq. \ref{Lambert}.} \label{Fig4}
\end{figure}

These fits are reproduced in Fig. \ref{Fig4} to illustrate how closely the limiting form Eq. \ref{Lambert} agrees with experiment, even for small values of $\sigma$. In the Figure, 
curves of different $\sigma$ are compared on a scale of reduced time $t_0 = t/ (1+\sigma)$. The data of Roinel {\it et al.} ($\sigma = 17$) are therefore consistent with a function whose Fourier transform closely approximates $\omega^{-3/2}$ behaviour over more than two decades.  

The second case to consider is the application of a magnetic field that oscillates around zero, $H= h e^{i \omega t}$. The question can be posed, can the complex susceptibility predicted here actually be measured by ac-magnetometry?  

In order to address this question, it is necessary to `unpack' the Faughnan-Strandberg equations somewhat. Clearly, the mathematical analysis assumes that both $\sigma$ and $\tau_{\rm sl}$ (the latter of which which sets the units of time in the equations) remain constant as $M^0$ varies. Ostensibly, this is only possible if the field, and hence the energy spacing of the two levels (measured by frequency $\nu\propto h$) and the parameters $\sigma$ and $\rho$, remain fixed to a good approximation.  This, of course, applies to a step function field applied, but contradicts a field oscillating around zero. Nevetherless, the following relations can be established: 
$$
\tau_{\rm sl} = \frac{1}{2 \rho A (p^0+1/2)} = -\frac{n^0}{2 \rho A N}, 
$$
$$
\sigma = -\frac{n^0 A\tau_{\rm ph}}{\Delta \nu}. 
$$
So, for the equations to have meaning in the case of an oscillating applied field, we must have $\rho A \propto n^0 \propto \Delta \nu/A\tau_{\rm ph}$. Suppose that $\rho$ is fixed to its total, or average value, during the field cycle, $\rho\rightarrow \bar{\rho}$ and, for consistency with this, $\Delta \nu/\nu = {\rm constant}$. Then, given that $n^0\propto h$ and the average $\nu \propto h$, it follows that $A\propto \nu \propto 1/\tau_{\rm ph}$. These relations, i.e. lifetimes increasing at low energy, do not seem unreasonable. Hence, while it remains far from proved, it does seem plausible that an ac-susceptibility experiment could reveal the $1/\omega^{3/2}$ behaviour predicted here. 

The third, and final, case to consider is when a small oscillating field is applied to a system at equilibrium in a relatively strong applied field. In this case, the term $M/M^0$ multiplying $\sigma$ in Eq. \ref{meqreduced} can be set to unity, and the relaxation is exponential~\cite{mistake}.

\section{Discussion and Conclusions}\label{Conclusions}

To expand upon the remarks in the Introduction, an isolated system will always tend to exhibit non-exponential decay as energy and information are retained in the system, leading to a greater likelihood of `memory'. Conversely, connection to a heat bath, will generally suppress memory as energy and information are transferred to the bath and effectively lost to the system, leading to exponential relaxation. The phonon bottleneck, considered here,  demonstrates this latter point rather transparently. If the phonons remain at thermal equilibrium then the phonon population does {\it not} depend on the history of the relaxation because it is fixed to its equilibrium value. Hence there is no `feedback' to the magnetization and the decay is exponential. If, on the other hand, the phonon occupation number differs from its equilibrium value, then its decay towards equilibrium depends on the history of the decay of the spin populations: this information feeds back to affect the decay of those populations, leading to non-exponential relaxation. 
 
The value of $\alpha =3/2$ found here, loosely equivalent to stretched exponential ($e^{-t^\beta}$) relaxation with $\beta = 2-\alpha =  1/2$, may be compared to the values found in other magnetic systems. Ising spin glasses~\cite{Gunnarsson} display values of $\beta \approx 0.25-0.3$. In single molecule magnets, a value of $\beta = 1/2$ has been observed at millikelvin temperatures at resonant tunneling points and interpreted in terms of fluctuating random fields~\cite{Sangregorio, Thomas}. In spin ice, $\alpha$ evolves from $2$ down to about 1.5 at subkelvin temperatures~\cite{Billington}, where the value of roughly 1.5 has been associated with the diffusion of emergent magnetic monopoles in a spatially fractal environment~\cite{Hallen, Orendac}. It is perhaps unsurprising that that the few-body systems (paramagnets, single molecule magnets) seem to exhibit exponents that are rational fractions, while the many body systems (spin glasses, spin ice) exhibit more general exponent values.  

Among non-magnetic systems, there is much interest in the related problem of noise in two-level systems. One example is flux noise in SQUIDS~\cite{aquino}. Another is the fluorescence noise of quantum dots where values of $\alpha$ and high-frequency crossovers rather similar to those found in the present work have been observed~\cite{Pelton, Barkai}. In such systems, the anomalous noise originates in the connection of the two-level system to other energy states, either extrinsic~\cite{aquino} or intrinsic~\cite{Pelton, Barkai}, creating a certain analogy with the phonon bottleneck system. In this context, it would be interesting to study the magnetic noise directly in the case of the phonon bottleneck.

It is finally worth noting some general insights afforded by the present results.  Exponential relaxation indicates that the rate of some relaxing quantity is proportional to its displacement from equilibrium. In the present case $dv/dt$ is instead linear with $u$, the displacement of the phonon equilibrium (Eq. \ref{reducevu}), but the latter is a highly nonlinear function of $v$, and this is the origin of the anomalous relaxation. A situation that is mathematically (though not physically) similar arises in the case of $e^{-\sqrt{t}}$ relaxation in an ordered two-dimensional Ising model where the radius of relaxing circular domains (droplets) plays a similar role to $u$ in the phonon case~\cite{FisherHuse}. In physical terms, the example of the phonon bottleneck shows how just two relaxation processes, both in themselves exponential, may generate a $1/f^\alpha$ power spectrum relaxation, when sufficiently strongly coupled. The phonon bottleneck thus illustrates how anomalous relaxation can emerge in a basic, but realistic, model of a far from equilibrium, and essentially quantum mechanical, steady state. Given its simplicity, the model might lend itself to generalisations, and given current interest in two-level systems, either as qubits~\cite{Burkard} or as sources of noise~\cite{Muller}, it might find contemporary relevance. 

\acknowledgments{It is a pleasure to thank P. Holdsworth and S. Giblin for useful comments.}


\begin{thebibliography}{99}
\bibitem{power} For a driven system $2|\chi''(\omega)|/\omega\chi_T = F(\omega)$, the spectral shape function of the relaxation. If the fluctuation-dissipation theorem also applies then this quantity is proportional to the `power spectrum' for spontaneous fluctuations. In the present case, the fluctuation-dissipation theorem might reasonably be expected to apply in the `Onsager' sense that spontaneous fluctuations have only the same channels of relaxation as those prepared by application of a field, but the loss of definition of system temperature might complicate this and there do not seem to have been explicit studies of this problem.
\bibitem{mistake} A previous ArXiv version of this paper incorrectly implied that the non-exponential relaxation extends to this case.
\end{thebibliography}
\end{document}